\documentclass[aps,pra,showpacs,groupedaddress,twocolumn]{revtex4}
\usepackage{amsfonts}
\usepackage{amsmath}
\usepackage{mathrsfs}
\usepackage{amssymb}
\usepackage[dvips]{graphics,color}
\usepackage{graphicx}
\usepackage{dcolumn}
\usepackage{bm}
\begin{document}
\author{}
\bigskip%
\author{Hui Jing$^{1,2}$, Yuangang Deng$^1 $, and Weiping Zhang$^2 $}
\affiliation{$^1$Department of Physics, Henan Normal University,
Xinxiang 453007, China \\
$^2$State Key Laboratory of Precision Spectroscopy, Department of
Physics, East China Normal University, Shanghai 200062, People's
Republic of China}
\title
{Quantum Control  of Light through an Atom-Molecule Dark State}
\begin{abstract}
We propose to use a quantized version of coherent two-color
photoassociation to realize a hybrid device for quantum control of
light. The dynamical features of this system are exhibited,
including the slowing down or storage of light and the molecular
matter-wave solitons. This may indicate a hybrid atom-molecule
quantum device for storage and retrieve of optical information.
\end{abstract}
\date{\today}
\pacs{42.50.-p, 03.75.Pp, 03.70.+k} \maketitle
%EndExpansion

\section{Introduction}

The experimental realization of Bose-Einstein condensates (BECs) in
ultracold atomic gases has led to tremendous advances from
traditional atomic, molecular, and optical (AMO) physics \cite{Atom
Optics} to current quantum information science \cite{computation,
teleportation}. Recently, an intriguing atom-molecule dark state was
observed in coherent two-color photoassociation (PA) \cite{quasi-one
2}, which has been considered as an efficient way to achieve higher
production rates of molecules \cite{two-photon detuning} from
ultracold atoms. In view of their internal properties and long-range
anisotropic interactions \cite{long-range}, the assembly of
heteronuclear molecules \cite{heteronuclear molecule, mixture3, h-m
0, h-m 1, f-f mixture} have also been actively pursued with various
important applications \cite{quantum 1, quantum 2, quantum 3,
precision measurements,electric dipole moment}, such as a polar
molecular quantum computer \cite{quantum 1, polar molecules BEC}. In
the light of these developments it is timely to investigate the
method of encoding and manipulating quantum optical state through
the atom-molecule dark state. Such processes will provide new
insights on current efforts of optical PA or quantum superchemistry
with the goal of designing a hybrid atom-molecule device for quantum
control of photonic information.

In this work we study such a scenario by transferring the quantum
state of an associating light to an atom-heternuclear molecule dark
state \cite{quasi-one 2, two-photon detuning}. This allows us to
study the effects of initial populations imbalance on the optical
storage process. In particular, our work compares the results for
atom-molecule systems with the more familiar light-storage schemes
in atomic samples \cite{slow light 0}. For a given number of atoms,
the signal light is slowed more in the atom-molecule hybrid system,
indicating some advantages over atomic slow-light media. Hence our
present proposal, together with e.g. a cascaded molecular
transition, may indicate a hybrid device for optical storage,
processing, and retrieval.

\begin{figure}[ht]
\includegraphics[width=0.7\columnwidth]{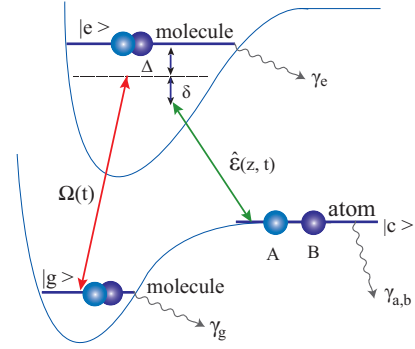}
\caption{(color online). Heteronuclear molecular creation in an
initial two-species atomic Bose condensate via coherent two-color
PA, with free-quasi-bound-bound transition induced by a quantum
signal light and a classical coupling field. }
\label{fig1}\end{figure}

\section{Model and Solutions}

As Fig. 1 illustrated, the initial ultracold bosonic two-species
atomic condensates (with populations $N_a$ or $N_b$) are
photoassociated into the excited molecular state $|e\rangle$ by a
quantized signal light, which is then dumped into the ground
molecular state $|g\rangle$ by another classical coupling light. The
signal pulse is described by the dimensionless operator
\begin{eqnarray}\label{eqn:field}
\hat{E}(z, t)=\sqrt{\frac{\hbar\nu}{2\epsilon_0 L}}\hat{\cal E}(z,
t)\exp(i\frac{\nu}{c}(z-ct)),
\end{eqnarray}
where $L$ is the quantization length in the $z$ direction, $\nu$ is
the PA light frequency and $\hat{\cal E}(z, t)$ is the slowly
varying amplitude. We focus on the role of coherent couplings of
photons and matter waves by ignoring the collisions of a dilute or
Feshbach-resonance-tuned medium \cite{FR method}. This is a safe
approximation for the short lifetime of associated dimers
\cite{quasi-one 2}. The operators of signal light and matter waves
satisfy the commutation relations, $[\hat{E} (z, t),\ \ \hat{E}
^\dagger (z',t)] =\frac{\nu}{\epsilon_0}\delta (z-z')$,
$[\hat{\phi_{i}}(z, t), \hat{\phi_{j}}^{\dagger}(z', t)]=\delta_{ij}
\delta (z-z'),$ respectively. The dynamics of this system is
described in the simplest level by the interaction Hamiltonian
($\hbar=1$)
\begin{align}\label{eqn:Hamiltonian1a}
\hat{\mathcal{H}}=&\Delta\int dz\hat{\phi_{e}}^{\dagger} (z, t)
\hat{\phi_{e}}(z, t)+\delta\int dz\hat{\phi_{a}}^{\dagger} (z, t)
\hat{\phi_{a}}(z, t)\nonumber\\
&-\int dz\bigr[g \hat{\cal E}(z, t)\hat{\phi_{e}}^{\dagger} (z,
t)\hat{\phi_{a}}(z, t)\hat{\phi_{b}}(z, t)\nonumber\\
&+\Omega \hat{\phi_{e}}^{\dagger} (z, t) \hat{\phi_{g}}(z,
t)+H.c.\bigr],
\end{align}
where $\Delta$ or $\delta$ is the one- or two-photon detuning,
$\Omega$ is the Rabi frequency of the coupling field, and
$g=\wp\sqrt{\frac{\nu}{2\hbar\epsilon_0 V}}$ is the photon-matter
waves coupling coefficient with $\wp$ being the transition-dipole
moment of $|c\rangle-|e\rangle$ transition by $\hat{E}(z, t)$
\cite{slow light 0}. Without loss of generality, we assume that the
signal field amplitude $\hat{\cal E}$ and control field amplitude
$\Omega$ are real whose   phase factor can be absorbed by a global
gauge transformation of the field operators \cite{two-photon
detuning}. Here we first drop off the usual kinetic and the trapping
terms by considering a uniform system and the effects due to these
terms will be discussed later.

With the slowly varying amplitude approximation \cite{slow light 0},
the propagation equation of the signal light can be written as
\begin{eqnarray}\label{eqn:light1}
(\frac{\partial}{\partial t}+c\frac{\partial}{\partial
z})\hat{\cal E}(z, t)= i gL\hat\phi_a^{\dagger} (z,
t)\hat\phi_b^{\dagger} (z, t)\hat\phi_e (z, t).
\end{eqnarray}
Meanwhile, the evolutions of atomic field operators are described by
the following Heisenberg equations
\begin{subequations}
\begin{eqnarray}\label{eqn:field1}
\dot{\hat{\phi_a}} &=& -i\delta\hat\phi_a- \gamma_{a}
\hat{\phi_{a}}+i g \hat{\mathcal {E}}^{\dagger}
\hat{\phi_{b}}^{\dagger} \hat{\phi_{e}},\\
\dot{\hat{\phi_b}} &=& - \gamma_{b} \hat{\phi_{b}}+i g \hat{\mathcal
{E}}^{\dagger}
\hat{\phi_a}^{\dagger} \hat{\phi_e},\\
\dot{\hat{\phi_e}} &=& -i \Delta \hat{\phi_e}-\gamma_{e}
\hat{\phi_e}+ i g \hat{\cal E} \hat{\phi_a} \hat{\phi_b}+i
\Omega \hat{\phi_g},\\
\dot{\hat{\phi_g}} &=& - \gamma_{g} \hat{\phi_g}+i \Omega
\hat{\phi_e},
\end{eqnarray}
\end{subequations}
where $\gamma_{a}$, $\gamma_{b}$,\ $\gamma_{e}$ and $\gamma_{g}$
denote the decay rates of corresponding matter-wave states. In order
to obtain a closed-form signal-light propagation equation, it is a
key step to study the evolutions of the following hybrid operators,
\begin{eqnarray}\label{eqn:field00}
\frac{\partial}{\partial t}({\hat{\phi_{a}}}^{\dagger
}{\hat{\phi_{b}}}^{\dagger } \hat{\phi_{e}}) &=& i{g} \hat{\cal
E}{\hat{\phi_{a}}}^{\dagger
}{\hat{\phi_{a}}}{\hat{\phi_{b}}}^{\dagger
}{\hat{\phi_{b}}}+i\Omega{\hat{\phi_{a}}}^{\dagger
}{\hat{\phi_{b}}}^{\dagger } \hat{\phi_{g}} \nonumber\\&& -
(\gamma_2-i\Delta-i\delta){\hat{\phi_{a}}}^{\dagger}
{\hat{\phi_{b}}}^{\dagger}\hat{\phi_{e}} \nonumber\\
&&-i{g}\hat{\cal E}({\hat{\phi_{a}}}^{\dagger}
\hat{\phi_{a}}+{\hat{\phi_{b}}}^{\dagger}
\hat{\phi_{b}}){\hat{\phi_{e}}}^{\dagger} \hat{\phi_{e}},
\end{eqnarray}
\begin{eqnarray}\label{eqn:field11}
\frac{\partial}{\partial t}(\hat\phi_a^\dagger
\hat\phi_b^\dagger\hat\phi_g) &=&
-(\gamma_1-i\delta)\hat\phi_a^{\dagger}\hat\phi_b^{\dagger}
\hat\phi_g +i\Omega\hat\phi_a^\dagger \hat\phi_b^\dagger\hat\phi_e\nonumber\\
&& - i{g}\hat {\cal{E}}(\hat\phi_a^\dagger
\hat\phi_a+\hat\phi_b^\dagger \hat\phi_b)\hat\phi_e^\dagger
\hat\phi_g,
\end{eqnarray}
%%%%%%%%%%%%%%%%%%%%%%%%%%%%%%%%%%%%%%%%%%%%%%%%%%%%%%%%%%%%%%%%%%%%%
with the transversal decay rates $\gamma_{1}=\gamma_{a} +\gamma_{b}
+ \gamma_{g}$ and $\gamma_{2}=\gamma_{a} +\gamma_{b}+\gamma_{e}$.
These equations can be rewritten as
\begin{eqnarray}\label{eqn:field2}
\hat\phi_a^\dagger \hat\phi_b^\dagger\hat\phi_e &=&
-\frac{i}{\Omega}\frac{\partial}{\partial t}(\hat\phi_a^\dagger
\hat\phi_b^\dagger\hat\phi_g) -\frac{i}{\Omega}
(\gamma_1-i\delta)\hat\phi_a^{\dagger}\hat\phi_b^{\dagger}
\hat\phi_g \nonumber\\
&& + \frac{g}{\Omega} \hat {\cal{E}}(\hat\phi_a^\dagger
\hat\phi_a+\hat\phi_b^\dagger \hat\phi_b)\hat\phi_e^\dagger
\hat\phi_g,
\end{eqnarray}
\begin{eqnarray}\label{eqn:field3}
{\hat{\phi_{a}}}^{\dagger }{\hat{\phi_{b}}}^{\dagger }
\hat{\phi_{g}} &=& -\frac{i}{\Omega}\frac{\partial}{\partial
t}({\hat{\phi_{a}}}^{\dagger }{\hat{\phi_{b}}}^{\dagger }
\hat{\phi_{e}})-\frac{g}{\Omega} \hat{\cal
E}{\hat{\phi_{a}}}^{\dagger
}{\hat{\phi_{a}}}{\hat{\phi_{b}}}^{\dagger }{\hat{\phi_{b}}}
\nonumber\\&& -\frac{i}{\Omega}
(\gamma_2-i\Delta-i\delta){\hat{\phi_{a}}}^{\dagger}
{\hat{\phi_{b}}}^{\dagger}\hat{\phi_{e}} \nonumber\\
&&+\frac{g}{\Omega}\hat{\cal E}({\hat{\phi_{a}}}^{\dagger}
\hat{\phi_{a}}+{\hat{\phi_{b}}}^{\dagger}
\hat{\phi_{b}}){\hat{\phi_{e}}}^{\dagger} \hat{\phi_{e}}.
\end{eqnarray}

It should be noted that Eq. (\ref{eqn:field2}) and Eq.
(\ref{eqn:field3}) can be greatly simplified under the weak
excitation approximation (WEA): the control field is much stronger
than the signal light at all times and thus the density of signal
photons can be taken as much less than that of atoms. This means
that only a small ratio of atoms are converted into molecules, which
is the case in the recent two-color PA experiment \cite{quasi-one
2}. With the WEA at hand, after some algebra we find in the lowest
non-vanishing order
\begin{eqnarray}\label{eqn:weak2}
{\hat{\phi_a}}^{\dag }{\hat{\phi_b}}^{\dag } \hat{\phi_g} &\approx&
- \frac{g \hat{\cal E}}{\Omega} {\hat\phi_a}^\dagger\
{\hat\phi_a}\hat{\phi_b}^\dagger\ \hat\phi_b.
\end{eqnarray}
Hence Eq. (\ref{eqn:field2}) can be rewritten as
\begin{align}\label{eqn:weak3}
{\hat{\phi_a}}^{\dag }{\hat{\phi_b}}^{\dag } \hat{\phi_e}\approx
i\frac{g N_a N_b}{\Omega^{2}} \frac{\partial}{\partial t}\hat{\cal
E}(z, t)\!&-\! i \frac{g N_a N_b}{\Omega^{3}}
\frac{\partial\Omega(t)}{\partial t}\hat{\cal E}(z, t),
\end{align}
%\end{subequations}
where $N_{a,b}=\hat{\phi}_{a,b}^{\dagger }\hat{\phi}_{a,b}$ is the
population of atoms A or B, which can be assumed as constant in the
WEA. Substituting Eq. (\ref{eqn:weak3}) into Eq. (\ref{eqn:light1})
yields
\begin{eqnarray}\label{eqn:light2}
(\frac{\partial}{\partial t} &+& \frac{c}{1+\frac{g^{2}LN_a
N_b}{\Omega^{2}}}\frac{\partial}{\partial
z})\hat{\cal E}(z,t)= \nonumber\\
&&= \frac{1}{1+\frac{g^{2}LN_a N_b}{\Omega^{2}}}\frac{g^{2}LN_a
N_b}{\Omega^{3}} \frac{\partial\Omega}{\partial t} \hat{\cal E}(z,
t).
\end{eqnarray}
Clearly, for a time-independent coupling field, we have a steady
group velocity of the signal, and the temporal profile or the
spectrum of the signal pulse remains unchanged during its slowing
down process, just as in a three-level atomic ensemble \cite{slow
light 0}. For a time-dependent coupling field, however, the
rand-hand side of Eq. (\ref{eqn:light2}) leads to an adiabatic Raman
enhancement of the signal pulse
\begin{eqnarray}\label{eqn:light3}
\hat{\mathcal {E}}(z,t)=\frac{\cos\theta(t)}{\cos\theta(0)}\
\hat{\mathcal {E}}(z-\int_0^tv_gdt',0),
\end{eqnarray}
where $v_{g} = c \cos^{2}\theta$ is the group velocity of the signal
light and $\theta$ is the mixing angle between light and matter-wave
components, i.e.,
\begin{align}\label{eqn:group1}
\tan^2\theta&={g^2N_a N_bL}/{\Omega^2(t)},\nonumber\\
v_{g}^{-1} &= \left(1+\frac{\tilde{g}^{2}N_a
N_b}{\Omega^{2}}\right)/c,
\end{align}
with $\tilde{g}=g\sqrt{L}$. Obviously, if the classical field is
adiabatically turned off by rotating the mixing angle $\theta$ for
$\pi/2$, the signal light will be fully stopped within the medium or
in the created atom-molecule dark state [4].

For the atomic slow-light medium \cite{slow light 0}, the group
velocity of signal light is:
$v_g=c/\bigl(1+\frac{{g}^2N}{\Omega^2})$, i.e., being proportional
to $\sim N^{-1}$, where $N$ can be regarded as the number of initial
trapped atoms in the WEA; however, in our situation, this velocity
is proportional to $\sim N^{-2}$ (for an initial balanced sample
$N_a=N_b=N/2$). Hence the technique of atom-molecule dark state may
have some advantage over the scheme of purely atomic spin waves
since the signal light can be much slowed down by starting from the
same total number of atoms.

The quantum state transfer process is also observed through the form
of the closed-channel molecular field. From Eqs. (7-9), it is
straightforward to find
\begin{align}
\hat{\phi_{g}}(z,t)\approx
k\hat{\mathcal {E}}(z-\int_0^tv_gdt',0), \nonumber\\
k=- \frac{g\sqrt{N_a N_b}}
{\Omega(0)}\sqrt{\frac{\Omega^2(0)+\tilde{g}^2N_a N_b}
{\Omega^2(t)+\tilde{g}^2N_a N_b}},
\end{align}
%%%%%%%%%%%%%%%%%%%%%%%%%%%%%
Obviously, for the initial stage ($\Omega^2\gg \tilde{g}^2N_a N_b$),
we have a purely photonic state, i.e., $\theta(0)=0$ or
$$\hat\phi_g(z,0)=0 ,$$
but when the coupling light is shut down adiabatically, the quantum
state of the associating light is fully encoded into the created
molecules via the mapping
\begin{align}
\sqrt{L}\hat{\phi_{g}}(z,t)= -\hat{\mathcal
{E}}(z-\int_0^tv_gdt',0),
\end{align}
which thus indicates a possible quantum memory device based on
coherent two-color PA technique.

\section{Role of Populations Imbalance}

As Fig. 2 shows, the initial populations imbalance of the atoms A, B
($\eta=N_b/N_a$) also plays a role in the optical storage process,
and the optimal conversion is reached for the balanced case
($\eta=1$). In our calculations, the initial total atomic number is
taken as $N=N_a+N_b=3.0\times10^6$. In current experimental
conditions \cite{polar molecules experiment}, the strength of
coupling field can be chosen as (in the unit of MHz)
\begin{align}
\Omega(t)=&10\pi(1-0.5\tanh[0
.15(t-15)] \nonumber\\
&+0.5\tanh[0.15(t-125)]),
\end{align}
which guarantees that the group velocity can be adiabatically
reduced to zero in the photon-storage stage and then the signal
light can be re-accelerated in the retrieval stage. The control
field being similar to Eq. (16) is also used in the atomic
slow-light medium \cite{slow light 0}. For the other cases of
initial population imbalance, e.g., $\eta=15$ as in the very recent
experiment of creating polar molecules KRb \cite{heteronuclear
molecule}, some deviation from the optimized case can appear (see
Fig. 2).

%%%%%%%%%%%%%%%%%%%%%%%%%%%%%%%%%%%%%%%%%%%%%%%%%%%%%%%%%%%%%%%
\begin{figure}[ht]
\includegraphics[width=0.75\columnwidth]{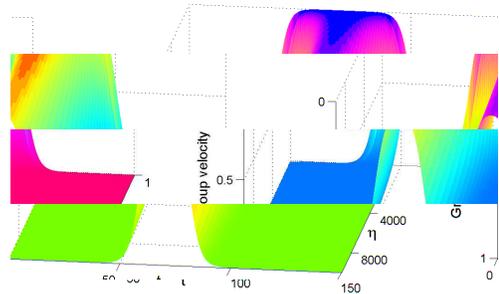}
\caption{(color online). The group velocity of signal light as a
function of time for different populations imbalances $\eta$. The
quantized length $L=1 mm$, the coupling coefficient $g=10$, the
velocity is scaled by $c$ and the time is in the unit of $\mu$s. }
\label{fig1}\end{figure}
%%%%%%%%%%%%%%%%%%%%%%%%%%%%%%%%%%%%%%%%%%%%%%%%%%%%

Fig. 3 shows the different features of slow light propagation in
four different kinds of matter-wave mediums: (i) the atomic
ensemble, (ii)-(iii) the assembly of homonuclear or heteronuclear
diatomic molecules [5, 21], and (iv) the assembly of heteronuclear
triatomic molecules ABC [5] (for the trimer case, the calculation of
the group velocity is completely similar to the above dimer cases).
In addition, we see from Fig. 3 that, by choosing a higher initial
atomic populations, the optical storage process can be significantly
improved.

\begin{figure}[ht]
\includegraphics[width=0.81\columnwidth]{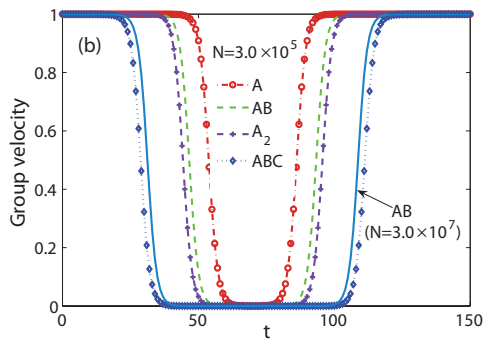}
\caption{(color online). The group velocity of the signal light as a
function of time in different matter-wave mediums, with the optimal
or balanced initial populations imbalance.} \label{fig1}\end{figure}

It is worth mentioning that in the above discussions we have ignored
the decay of molecular states. However, it is readily to show that,
after including these decay terms, the group velocity of the signal
light is still in the form of  Eq. (\ref{eqn:group1}) but with the
following substitution
%%%%%%%%%%%%%%%%%%%%%%%%%%%%%%%%%%%%%%%%%%%%%%%%%%%
$$\Omega \rightarrow \sqrt{\Omega^2+\gamma_1\gamma_2}.$$
%%%%%%%%%%%%%%%%%%%%%%%%%%%%%%%%%%%%%%%%%%%%%%%%%%%
Clearly, due to the decay terms, one may reach a $nonzero$ group
velocity even when the classical field is turned off. For the
typical parameters of the molecules KRb \cite{polar molecules
experiment}, we can take $N_\mathrm{K}=1.0\times10^6$,
$N_{\mathrm{Rb}}=5.0\times10^6$, $\tilde{g}\approx50$ s$^{-1}$, and
%%%%%%%%%%%%%%%%%%%%%%%%%%%%%%%%%%%%%%%%%%%%%%%%%%%%%%%%%%
$$\gamma_1=2\pi\times97Hz, \gamma_2=2\pi\times5.7MHz.$$
%%%%%%%%%%%%%%%%%%%%%%%%%%%%%%%%%%%%%%%%%%%%%%%%%%%%%%%%%%%
The velocity limit can be estimated as $v^c_g\approx0.524$
km$\cdot$s$^{-1}$. In particular, for a sufficient state transfer,
the PA time duration should satisfy $t\ll \gamma_{1}^{-1}\sim1.6$ ms
\cite{quantum memory, quantum memory 1}, which can be fulfilled in
current experiments \cite{quantum memory experiment 1, quantum
memory experiment 2}.

We also note that some optimized methods exist for a maximum
efficiency of optical storage and retrieval, such as the recent
works of Novikova {\it et~al.} by using an optimized signal pulse
shape in an atomic medium \cite{Optimal 1,Optimal 2}. For the
present atom-molecule system, in order to avoid incoherent
absorptive loss, the frequency components of the signal pulse must
fit well with the slow-light spectral window, i.e., $t_s^{-1}\ll
\sqrt{d}v_g/L\sim N^{-1} $, where $t_s$ is the temporal length of
the signal pulse, $d=g^2N_aN_bL/(\gamma c)$ is optical depth of the
medium. Thereby, to avoid ``leakage" of the pulse outside the
medium, the following condition should be fulfilled
\begin{equation}
v_gt_s\sim N^{-1}\ll L.\nonumber
\end{equation}which means that $v_g$ must be small
enough for the entire signal pulse to be spatially compressed into
the medium, with also a large optical depth d. For contrast, by
using a purely atomic medium, we have $d=g^2NL/(\gamma c)$
\cite{Optimal 1} and thus $v_gt_s\sim N^{-{1}/{2}}$ \cite{Optimal
2}. Obviously, for the same initial total atomic number, the
quantum light storage using the atom-molecule dark state may have
some advantage over the familiar atomic spin-wave scheme.

\section{Molecular Soliton Laser}

Now we show that, by taking into account of the particle collisions
in the quantum state transfer process, it is also possible to
realize a molecular matter-wave soliton laser. To this end, we
consider coherent atom-molecule conversion process which is
described by the following total Hamiltonian
\begin{eqnarray}
\hat{H}_{tot}=\hat{H}_0+\hat{H}_{coll}+\hat{\mathcal{H}},
\end{eqnarray}
where $\hat{\mathcal{H}}$, as in Eq. (2), denotes the coherent
free-quasi-bound-bound transition, $\hat{H}_0$ and $\hat{H}_{coll}$
describe the free motions and the particle collisions, respectively,
%%%%%%%%%%%%%%%%%%%%%%%%%%%%%%%%%%%%%%%%%%%%%%%%%%%%%%%%%%%%%%%%%
\begin{align}
&H_0=\sum_{i}\int
dz\hat\phi_i^{\dag}(-\frac{1}{2m_i}\frac{\partial^2}{\partial
z^2}+ V_i)\hat\phi_i,\nonumber\\
&H_{coll}=\sum_{i,j} U_{ij}\int dz\hat\phi^{\dag}_i
\hat\phi^{\dag}_j\hat\phi_j\hat\phi_i, \end{align}
%%%%%%%%%%%%%%%%%%%%%%%%%%%%%%%%%%%%%%%%%%%%%%%%%%%%%%%%%%%%%%%%
were $V_i(z)(i, j=a, b, g) $ denotes the longitudinal external
effective potential and one can choose $V_a(z)$=$V_b(z)=0$ in the
following derivation \cite{slow light 0}, $m_i$ is the particle
mass, $U_{ij}$ denotes the $s$-wave scattering collisions between
the particles \cite{scattering length}. Then the evolution of
molecular matter-wave field is written as
\begin{eqnarray}\label{eqn:2}
i\frac{\partial\hat\phi_g}{\partial
t}&=&[-\frac{1}{2(m_a+m_b)}\frac{\partial^2}{\partial z^2}
+U_{ab}\hat\phi_a^{\dag}\hat\phi_b  \nonumber\\
&+&U_{gg}\hat\phi_g^{\dag}\hat\phi_g+ V_g(z)]\hat\phi_g
-\Omega\hat\phi_e.
\end{eqnarray}

\begin{figure}[ht]
\includegraphics[width=0.83\columnwidth]{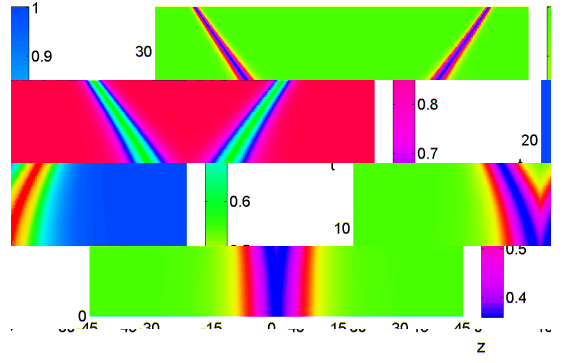}
\caption{(color online). Splitting grey solitons under the influence
of background amplitude decreasing, for $q=0.8$. The units of time
and trap size are in $\mu$s and $\mu$m, respectively.  }
\label{fig4}\end{figure}

For simplicity, we consider an initially trapped atomic ensemble and
also introduce the mean-field approach by replacing the operators by
the $c$-numbers, i.e., $\hat\phi_i(z,t)\rightarrow\Phi_i(z,t)$.
Thereby, for the closed-channel molecules, i.e., with the mixing
angle $\theta=\pi/2$, we obtain the nonlinear mean-field
Gross-Pitaevskii equation
\begin{eqnarray}\label{eqn:8}
i\frac{\partial\Phi_g}{\partial
t}\!=\!-\frac{1}{2(m_a+m_b)}\frac{\partial^2\Phi_g }{\partial z^2}
\!+\!V_{eff}\Phi_g\!+\!U_{gg}|\Phi_g^0|^2\Phi_g,
\end{eqnarray}
where the effective potential $V_{eff}=V_g+\sqrt{N_aN_b}U_{ab}$ can
also be moved by suitably tuning the value of $V_g$ \cite{slow light
0}. For $a_{gg}>0$, due to some balance between the repulsive
molecular collisions and the molecular kinetic energy, the above
well-known nonlinear equation can support a gray-soliton solution
\cite{soliton 1, soliton 2}
\begin{eqnarray}\label{eqn:9}
\Phi_g\!=\!\Phi^0_g(z,t)\bigr\{i\sqrt{1-q^2}+q\tanh\bigl[\frac{q}{\sqrt{\alpha}}(z-z_0(t))\bigl]\bigr\},
\end{eqnarray}
with $\alpha=(\sqrt{4\pi a_{gg}}|\Phi_g^0|)^{-1}$. $V_g $ is chosen
such that $V_{eff}=0$, the slowly-varying background function is
%%%%%%%%%%%%%%%%%%%%%%%%%%%%%%%%%%%%%%%%%%%%%%%%%%%%%%%%%%%%%%%%%%%%%
\begin{eqnarray}
\Phi^0_g(z,t)=\langle\hat\phi_g\rangle\exp(-i\int^t_{t_0}U_{gg}|\langle\hat\phi_g\rangle|^2
dt'),
\end{eqnarray}
%%%%%%%%%%%%%%%%%%%%%%%%%%%%%%%%%%%%%%%%%%%%%%%%%%%%%%%%%%%%%%%%%%%%%%
and $z$ is the inside-trap position. In addition, the ``grayness"
parameter is
%%%%%%%%%%%%%%%%%%%%%%%%%%%%%%%%%%%%%%%%%%%%%%%%%%%%%%%%%%%%%%%%%%%%%%%
\begin{eqnarray}
q=\sqrt{1-({v_\nu}/{v_s})^2}\leq 1,
\end{eqnarray}
%%%%%%%%%%%%%%%%%%%%%%%%%%%%%%%%%%%%%%%%%%%%%%%%%%%%%%%%%%%%%%%%%%%%%%%
with the Bogoliubov sound speed \cite{speed of sound}
$$v_s=[U_{gg}|\Phi_g^0|^2/(m_a+m_b)]^{1/2},$$
and the dark-soliton speed $v_\nu=$\.{z}$_0(t)$, ${z}_0(t)$ being
the central position of the molecular matter-wave soliton. Clearly,
$q=1$ corresponds to a dark soliton with $100\%$ density depletion.
We see from Fig. 4 that, by applying two counter-propagating control
fields, a second-order molecular grey soliton ($q=0.8$) starting
from the same position ($z=0$) can split into two solitons
propagating at opposite directions.

\section{Conclusion}

In conclusion, we have studied the slow light and quantum optical
storage process in coherent two-color PA process by considering a
quantized associating light. By taking into account of the particle
collisions, one may also create the molecular matter-wave solitons.
This may indicate a hybrid atom-molecule quantum device for storage
and retrieve of optical information. It is straightforward to study
other interesting configurations, such as the light-molecule
entanglement by applying a non-classical PA light \cite{FR method,
squeeze 2}, or the quantum switch by considering a multi-level
atom-molecule system.

As far as we know, our work sets up the first link between the
research fields of quantum memory and coherent atom-molecule
conversion. Due to the rapid experimental advances in both two
fields, this atom-molecule system may be potentially useful for
designing a hybrid atom-molecule quantum device of optical storage,
processing, and retrieval. In the future, we plan to study the
quantum memory in a boson-fermion mixture [5, 8], the slow light in
the BEC-BCS crossover of a fermionic atomic sample \cite{f-f
mixture}, and the polarization rotation of slow light \cite{AMO}
with the Laguerre-Gaussian signal modes.

\acknowledgments H.J. and Y.D. are supported by the National Science
Foundation of China (Grant No. 10874041), the NCET, and the ECNU Key
Lab Open Fund. Weiping Zhang is supported by the National Science
Foundation of China (Grant No. 10588402 and No. 10474055), the
National Basic Research Program of China (Grant No. 2006CB921104),
the Science and Technology Commission of Shanghai Municipality
(Grant No. 06JC14026 and No. 05PJ14038), the Program of Shanghai
Subject Chief Scientist (Grant No. 08XD14017), the Program for
Changjiang Scholars and Innovative Research Team, Shanghai Leading
Academic Discipline Project (Grant No. B480), and the Research Fund
for the Doctoral Program of Higher Education (Grant No.
20040003101).

%%%%%%%%%%%%%%%%%%%%%%%%%%%%%%%%%%%%%%%%%%%%%

%\bigskip

%\noindent

\end{document}